\documentclass{emulateapj}

\usepackage[usenames,dvipsnames]{color}

\advance \voffset by -0.5cm\relax

\usepackage{ulem}

\usepackage{natbib}
\def\apjl{\rm ApJL}
\def\apjs{\rm ApJS}

\def\aap{\rm AAP}

\def\msun{{\rm ~M}_{\odot}}

\def\rsun{{\rm ~R}_{\odot}}

\def\zsun{{\rm ~Z}_{\odot}}

\newcommand{\gtorder}{\mathrel{\raise.3ex\hbox{$>$}\mkern-14mu
            \lower0.6ex\hbox{$\sim$}}}
\newcommand{\ltorder}{\mathrel{\raise.3ex\hbox{$<$}\mkern-14mu
            \lower0.6ex\hbox{$\sim$}}}

\newcommand{\code}[1]{\texttt{#1}}

\newcommand{\STRACK}{\code{StarTrack }}

\shorttitle{Massive black-hole binaries}
\shortauthors{}

\begin{document}

\title{The Formation and Gravitational-Wave Detection of\\Massive Stellar Black-Hole Binaries}

\author{
Krzysztof Belczynski\altaffilmark{1,2}, 
Alessandra Buonanno\altaffilmark{3}, 
Matteo Cantiello\altaffilmark{4}, 
Chris L. Fryer\altaffilmark{5}, 
Daniel E. Holz\altaffilmark{6,7}, 
Ilya Mandel\altaffilmark{8}, 
M. Coleman Miller\altaffilmark{9},
and Marek Walczak\altaffilmark{1}
}

\affil{
$^{1}$ Astronomical Observatory, Warsaw University, Al. Ujazdowskie 4, 00-478 Warsaw, Poland (kbelczyn@astrouw.edu.pl)\\
$^{2}$ Center for Gravitational Wave Astronomy, University of Texas at Brownsville, Brownsville, TX 78520, USA\\
$^{3}$ Maryland Center for Fundamental Physics and Joint Space-Science Institute, Department of Physics, University of Maryland, College Park, MD 20742, USA\\
$^{4}$ Kavli Institute for Theoretical Physics, University of California, Santa Barbara, Kohn Hall, CA 93106, USA\\
$^{5}$ Computational Computer Science Division, Los Alamos National Laboratory, Los Alamos, NM 87545, USA\\
$^{6}$ Enrico Fermi Institute, Department of Physics, and Kavli Institute for Cosmological Physics, University of Chicago, Chicago, IL 60637, USA\\
$^{7}$ Theoretical Division, Los Alamos National Laboratory, Los Alamos, NM 87545, USA\\
$^{8}$ School of Physics and Astronomy, University of Birmingham, Edgbaston, Birmingham, B15 2TT, UK\\
$^{9}$ Department of Astronomy and Joint Space-Science Institute University of Maryland, College Park, MD 20742-2421, USA\\
}

\begin{abstract}
If binaries consisting of two $\sim100~M_\odot$ black holes exist they would
 serve as extraordinarily powerful gravitational-wave sources, detectable to
 redshifts of $z\sim2$ with the advanced LIGO/Virgo ground-based detectors.
 Large uncertainties about the evolution of massive stars preclude definitive
 rate predictions for mergers of these massive black holes. We show that rates
 as high as hundreds of detections per year, or as low as no detections
 whatsoever, are both possible. It was thought that the only way to produce
 these massive binaries was via dynamical interactions in dense stellar
 systems. This view has been challenged by the recent discovery of several
 $\gtrsim150~M_\odot$ stars in the R136 region of the Large Magellanic Cloud.
 Current models predict that when stars of this mass leave the main sequence,
 their expansion is insufficient to allow common envelope evolution to
 efficiently reduce the orbital separation. The resulting black-hole--black-hole binary remains
 too wide to be able to coalesce within a Hubble time. If this assessment is
 correct, isolated very massive binaries do not evolve to be gravitational-wave
 sources. However, other formation channels exist. For example, the high
 multiplicity of massive stars, and their common formation in relatively
 dense stellar associations, opens up dynamical channels for massive black
 hole mergers (e.g., via Kozai cycles or repeated binary-single interactions).
 We identify key physical factors that shape the population of very massive
 black-hole--black-hole binaries. Advanced gravitational-wave detectors will provide important
 constraints on the formation and evolution of very massive stars.

\end{abstract}


\section{Introduction}

In the early universe massive ($M\gtrsim100\,M_\odot$) black holes (BHs) are believed to
form from the collapse of massive stars \citep{Fryer01}, and these BHs may be the seeds of the
supermassive BHs at the nuclei of galaxies \citep{Madau01,Whalen12}. 
These massive BHs have also been invoked to
explain ultraluminous X-ray sources (so termed because they emit at
10--100 times the Eddington rate for a $10\,M_\odot$ BH) in the
nearby universe \citep{colbert99,colbert04}. Until recently, both observations
of stellar clusters, e.g., \citet{Figer05} (although see also \cite{Massey03}),
and some theoretical arguments \citep{Mckee07} have suggested that stars above
$150\,M_\odot$ do not form at non-zero metallicities.  Including the
effects of mass loss from winds, even at 1/10th solar metallicity,
this assumption produces masses of BH systems in the nearby
universe in the tens of solar masses \citep{2010ApJ...714.1217B}.

However, the discovery of several stars with current masses greater
than $150\,M_\odot$ and initial masses up to $\sim 300\msun$ in the
R136 region of the Large Magellanic Cloud \citep{2010MNRAS.408..731C}
requires a rethinking of this argument.  There are at least some
environments in which stellar
masses can apparently extend well beyond $150\,M_\odot$, and if these
stars do not have extremely large wind loss rates, then after their
cores collapse they may leave behind BHs with masses in
excess of $100\,M_\odot$.  

It has been suggested that stars with initial masses between roughly
$150~M_\odot$ and $300~M_\odot$ at low to moderate metallicities explode due to an instability produced when
the core gets so hot that it produces electron/positron pairs, resulting in a
loss of energy which reduces the pressure and causes the core
to contract.  As this happens the nuclear burning accelerates and
can disrupt the star entirely if the star is unable to stabilize
itself.  If some fraction of superluminous supernovae are in fact
pair-instability supernovae \citep{woosley07,galyam09,cooke09,ofek14}
\citep[see, however,][]{Nicholl13}, then at least at metallicities 
around $\sim 0.1Z_\odot$ (where $Z_\odot=0.014$) massive stars above $150~M_\odot$ exist.

Gravitational waves (GWs) can potentially provide additional evidence for
the formation and evolution of these massive stars.  Massive stars are
usually found in binaries or multiple systems 
\citep[e.g.,][]{Kobulnicky07,Kobulnicky12,2012Sci...337..444S,2013A&A...550A.107S}
with mass ratios that are flat \citep{2013MNRAS.432L..26S}.  Indeed,
the most massive known binary has an estimated total mass of
$200$--$300\,M_\odot$ and a possible initial mass of $\sim
400\,M_\odot$ \citep{2013MNRAS.432L..26S}.  Thus massive BHs formed
from the evolution of these massive stars are likely to be partnered
with comparably massive BHs. If very massive stars (VMS) with initial masses 
above $150\,M_\odot$ also follow these trends, these stars may produce 
massive BH binaries, some of which may merge.  Keeping the binary's mass ratio fixed, 
the total energy emitted in GWs is proportional to the binary's total mass, so coalescing 
massive BH binaries can be detected much farther than stellar-mass BH binaries. Thus, they 
could be very significant sources for advanced GW detectors.

In this paper, we explore in detail the evolution of binaries with initial (zero-age main sequence) component masses above $150 M_\odot$ and the formation of very massive BH-BH binaries; the fate of binaries with initial component masses up to $150 M_\odot$ was explored elsewhere \citep{2012ApJ...759...52D}.  The
formation of merging, massive BH binaries depends sensitively
on a range of issues in stellar evolution --- for example the evolution of the core,
the expansion in the giant phase, and the details of the pair-instability supernovae.  
We discuss how these uncertainties lead to a wide range of predictions for the merger 
rates for these systems. Even an approximate measurement of the merger 
rates will place insightful constraints on these stellar processes.
We discuss these uncertainties in detail, calculating the full range 
of rate predictions for advanced LIGO/Virgo.

In Section~2 we discuss the physical processes and uncertainties
involved in the evolution of very massive binaries.  In Section~3 we
provide a range of predictions for the rates of formation and merger of
massive binary stellar-mass BHs, based on different
assumptions with different codes.  In Section~4 we consider dynamical
effects, particularly Kozai cycles and three-body interactions, and
find that even if most or all massive BH binaries are too
widely separated to merge on their own within a Hubble time, many will
be induced to merge by interactions with other objects.  In Section~5
we map these coalescence rate predictions for massive binaries to
predictions for detection rates in future GW observatories.  We
emphasize the importance of the merger and ringdown
phases of the gravitational waveform and cosmological effects, 
due to the high masses of the BHs and the
significant redshift out to which they can be observed.  Our discussion
and conclusions are in Section~6.

\section{BH-BH binary formation physics}

In this section we outline and discuss the basic evolutionary processes that 
are involved in the formation of close (with coalescence times below the 
Hubble time) BH-BH binaries from very massive stars. 
Three major sources of uncertainty are involved: 

\begin{enumerate} 

\item {\it Common envelope evolution:} During this stage one of the stars in the binary expands
such that its envelope surrounds both stars.  In order to ultimately
lead to a compact binary, the drag on the binary due to the envelope
must shrink the semimajor axis, but must eject the envelope before
the cores merge.  There are many uncertainties about the efficiency
with which this happens, as well as about the conditions necessary for
common envelope evolution as opposed to stable mass transfer via Roche 
lobe overflow.

\item {\it Stellar radius expansion:}  In order for
the common envelope phase to be effective, it is necessary that the
stars in the binary expand significantly when they go off the main
sequence, because otherwise the stars would have had to be nearly in
contact and would thus have likely merged due to tidal effects. However,
current codes suggest the emergence of new effects for very massive
stars (such as convection that encompasses nearly the entire star) that
may suppress such expansion.  These effects appear to have complex
dependencies on metallicity, rotation, and possibly even the code used
to do the analysis.

\item {\it Black-hole formation:} Massive stars 
(with initial mass $\lesssim 150\msun$) form an iron core and undergo core collapse at
the end of their evolution, thereby forming BHs. Very massive stars
($\lesssim 300\msun$) are
potentially unstable to pair creation well before iron core formation. The
subsequent pair-instability supernova may disrupt the entire star and thus 
prevent BH formation. For even more massive 
stars pair creation cannot overcome the gravity of the collapsing star, and it is 
expected that the entire star will end up forming a BH. The transition 
from one regime to another depends on the internal temperature structure of a star 
and is very sensitive to assumptions in the stellar evolutionary codes.
Additionally, there is significant uncertainty about the BH mass (whether 
it forms from a supernova explosion with its associated mass ejection, or whether 
the BH forms from the collapse of the entire star). Lastly,
it is possible that the resulting BH receives a significant natal kick,
disrupting the binary and eliminating massive BH-BH systems.

\end{enumerate}

We now discuss each of these issues in greater depth.

\subsection{Common envelope}

The formation of a close binary of two BHs from the initially wide 
binary of two massive stars, which are large objects with Zero Age Main
Sequence (ZAMS) radii in excess of $10$--$30\rsun$ \citep[e.g.,][]{2000MNRAS.315..543H}, 
requires a process that leads to significant and fast orbital 
decay. In isolated evolution, where there is no dynamical interaction 
with other stars, a common envelope (CE) phase in which the expanding envelope of 
a post-main-sequence star engulfs the entire binary system is the only available
process for sufficient orbital decay.  For a recent comprehensive review of the physics of CEs and the many 
remaining uncertainties see \citep{2013A&ARv..21...59I}.

To motivate this process in more detail, we recall that a star
undergoes a series of burning phases, where the core contracts and
heats up sufficiently such that the ashes of the previous phase are
ignited and burn.  Concurrent with these phases the
envelope can also undergo phases of expansion.  When the envelope
reaches the companion, the friction of the companion orbiting in the
envelope transfers orbital energy into the envelope. As a result, the
envelope may eventually become unbound, leaving a binary system
composed of the primary hot core and its mostly unaffected
companion. A large orbital contraction (factors of $\gtrsim 10$--$100$)
is expected for envelopes with mass $\gtrsim 10\%$ of the core mass 
($\gtrsim 10$--$100\msun$ for massive BHs). For binaries of two massive stars this process may
occur twice, as first the primary and then the secondary undergo
evolutionary expansion.  This picture is generally accepted but, as we
now describe, the details of the CE phase are both highly uncertain and
extremely important for predicting the properties of any resulting
stellar-mass BH binaries.

First, when the more massive star comes into contact with its Roche lobe, the 
binary is very likely eccentric and the individual stars are unlikely to have 
spins that are synchronized with the orbital motion because the evolution of 
massive stars is very rapid compared to the tidal timescale.  Mass transfer 
therefore occurs during periastron passage. Whether such intermittent mass transfer 
continues, or whether it leads rapidly to circularization/synchronization is not 
known. This translates into a highly uncertain initial separation at the onset of 
the CE phase.

Second, it is not fully understood which binary configurations lead to a CE (i.e., 
mass transfer proceeding on a dynamical timescale) and which lead to stable Roche 
lobe overflow (RLOF, i.e., mass transfer proceeding on the donor's nuclear or thermal 
timescale).  A standard argument begins by noting that the angular momentum of a 
circular binary with a semimajor axis $a$, total mass $M$, and reduced mass $\mu$ is 
$\mu\sqrt{GMa}$.  In conservative mass transfer, $M={\rm const}$. Transfer from the 
more to the less massive star increases $\mu$, whereas transfer from the less to the 
more massive star decreases $\mu$.  Angular momentum conservation thus implies that 
mass transfer from the more to the less massive object shrinks the binary. This is 
often used to argue that such transfer is unstable (compared to the stable mass 
transfer from the less to the more massive star).

The real situation is more complicated, particularly for very massive binaries. Mass 
and angular momentum can be lost from the system via winds, in some cases mass loss 
can shrink stars faster than the orbital separation shrinks, and it is in particular 
uncertain what happens for stars with radiative envelopes in large mass ratio 
binaries. Typically it is estimated that for mass ratios $q \lesssim 1/3$--$1/2$ (where $q$ 
is the ratio of the mass of the companion to the mass of the donor) and for donors 
with deep convective envelopes, RLOF will develop into CE.  Both outcomes, CE and 
stable RLOF, are adopted in the literature
\citep[e.g.,][]{Hjellming1987,Tauris1999,Wellstein2001,Ivanova2003,Dewi2003,2008ApJS..174..223B,   
Mennekens2014}.

Third, there is no self-consistent physical model to calculate reliably the post-CE 
orbital separation, which as we explained above is crucial to whether a system merges 
into one object or leads to a compact object binary, and whether such a binary can 
merge within the age of the universe.  There are two existing approaches. One uses an 
energy balance prescription which unfortunately does not conserve the angular 
momentum of the system \citep{Webbink1984}. The other uses an angular momentum balance prescription 
which unfortunately does not conserve energy \citep{Nelemans2005}. The two methods
may differ by as much as an order of magnitude in their predictions for post-CE orbital separation.  
Each method has had success in reproducing the orbital periods of separate groups of post-CE white 
dwarf binaries. It is not known which method (if either) is a reasonable approximation 
for very massive binaries with BHs.

Fourth, the donor envelope binding energy directly affects the fate of the binary after 
the CE phase, because as we describe above the envelope must be ejected to leave the 
core. If the binding energy is too large, the components will merge.  If the binary 
survives, the final separation depends on the value of the binding energy.  
There is, however, ambiguity in the definition of the envelope binding energy because 
it is difficult to precisely define the core-envelope boundary.  Different choices in the 
literature \citep[e.g.,][]{Tauris2001,Podsiadlowski2003,Voss2003,Xu2010,Loveridge} include 
a specific level of H-depletion, an entropy jump, or the position of the H-burning shell. 
Given that most of the mass and binding energy of the envelope are from the part of the envelope 
near the core, these choices may influence common envelope calculations. It has been suggested that 
double compact object merger rates may be affected by as much as an order of magnitude (Tauris, private 
communication 2012), while recently it was argued that the specific choice for the core-envelope 
boundary does not play a significant role for massive stars (i.e., black hole progenitors with 
$M_{\rm zams} \approx 70-100 \msun$; \citealt{Wong2014}). If the internal 
energy of the envelope is significant, this could make ejection easier. Similarly, if 
there are nuclear burning shells in the envelope they could affect the ejection process, 
although the sign of this effect is unclear.  Finally, even for a given envelope binding energy the ejection 
process will be affected by the fraction of the dissipated orbital energy which goes into the 
ionization and dissociation of atoms and molecules in the envelope rather than into bulk 
kinetic energy.

Fifth, for many massive stars the most significant radial expansion occurs right after 
the main sequence, as the rapid core adjustment after the end of core hydrogen burning 
causes the star to rapidly cross the Hertzsprung gap (HG). Hence it is expected that many 
massive binaries will initiate the CE phase during the HG.  However, it is not clear 
whether at this point the stars already have a well developed core-envelope structure. 
Some models indicate that the entropy profile is rather flat throughout the HG (i.e., 
it has similar structure to a main sequence star), which means that once the CE begins 
it will always end in a merger of both components. Other models suggest that if the core 
of the donor is exposed during this evolutionary phase (i.e., by the inspiraling companion), 
it may remain at its compact size and thus there is the potential for CE survival.

\subsection{Stellar radius evolution}

For a common envelope to develop, the radius of at least one of the stars must increase 
more rapidly than the Roche lobe radius.  Because the Roche lobe radius is 
$R_{\rm RL}\sim (m/M)^{1/3}a$, where $m$ is the mass of the donor, $M$ is the mass of the 
binary, and $a$ is the binary separation, this requirement is approximately equivalent to 
the requirement that the fractional expansion of the radius of one of the stars must exceed 
the fractional increase in the binary separation.  Very massive stars at non-zero metallicity 
(e.g., Population I stars with heavy element fractions $Z=0.002$, such as those
in the Small 
Magellanic Cloud) may lose half or more of their mass to stellar winds during the main 
sequence (e.g., \citealt{2013MNRAS.433.1114Y}). 
If mass is removed from a circular binary without changing the specific angular momentum 
$\sqrt{GMa}$, the binary's separation would double when half of the mass is lost. Thus the 
binary separation can increase by a factor of $\sim 2$ on the main sequence, and therefore 
to initiate a CE phase one of the binary components needs to expand by more than a factor 
of $2$.

However, as indicated above, if there is a CE phase during the main sequence then the 
shallow entropy gradient leads to full inspiral and merger.  Survival of the CE phase requires 
that it be initiated beyond the main sequence, when the core-envelope structure is well-developed 
and thus the envelope can be ejected while the core is unaffected.   The radius evolution of a 
very massive star is therefore critical to the fate of very massive binaries, but this evolution 
is currently uncertain because it depends, rather sensitively, on the treatment of convection, 
stellar winds, and rotational mixing in stellar models.

For example, models of $500~M_\odot$ stars done with the {\tt Geneva} 
code \citep{2013MNRAS.433.1114Y} show no radial expansion beyond the main sequence for metallicities $Z=0.014$ (solar) and $Z=0.006$, 
and only a mild expansion for $Z=0.002$ (from $\sim 50 \rsun$ at the middle of the main sequence to $\sim 110 \rsun$ 
after hydrogen core exhaustion). 
Given that, as indicated above, the orbit can expand by up to 
$\sim 30\%$ during this time, it is difficult for a binary to enter the CE phase and thus to 
produce a close massive BH-BH binary.

This expected behavior is qualitatively distinct from that of stars below $\sim 100\msun$, which 
are believed to expand significantly after the main sequence.  The difference for very massive 
stars is that they are thought to have large convective cores (up to 90\% of the stellar mass at 
the ZAMS for a star with an initial mass of $200\msun$, and up to 95\% of the stellar mass for a 
star with an initial mass of $500\msun$; see \citealt{2013MNRAS.433.1114Y}).  These cores stem 
from the large radiation pressure at such high masses, and mean that fresh
hydrogen is constantly being 
brought to the center and helium is mixed to the photosphere.  In the models this leads to strong 
Wolf-Rayet winds that can remove almost all of the hydrogen from the envelope,
resulting in highly helium-rich stars that experience little expansion.

The treatment of mixing induced by convection and rotation, along with the adopted wind mass loss 
rates, are the crucial factors in modeling the radial expansion of very massive stars. However both 
prescriptions are quite uncertain. Typical treatments are mostly based on theoretical estimates 
that are only weakly constrained by observations (especially in the case of rare very massive stars). 
Therefore, large uncertainties exist in the resulting radius evolution, leading
to outcomes anywhere between two extremes: (a)~no/small radial expansion, no/small envelope 
mass after the main sequence \citep{2013MNRAS.433.1114Y}, (b)~significant radial expansion, large envelope mass after the main 
sequence. Latter results were obtained using MESA calculations of very
massive stars, where a decisive effect in the radial extension of the
radiation-dominated regions of
very massive stars is envelope inflation \citep[see, e.g.,][]{Kato:1985,Ishii:1999,2006A&A...450..219P,2012A&A...538A..40G}.

\begin{figure}
\begin{center}
\includegraphics*[width=0.47\textwidth]{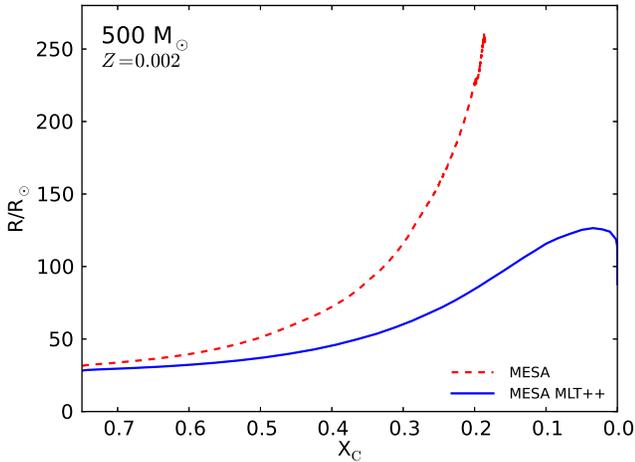}
\caption{Evolution of the photospheric radius as function of hydrogen core mass fraction 
(X$_{\rm C}$) for MESA calculations of a $500\msun$ model at $Z=0.002$. The dashed line shows a model 
where the convective energy transport 
is calculated according to the mixing length theory (MLT). The radial expansion is substantial and is 
associated with a density inversion in the stellar envelope. 
In this situation very short timesteps pose problems for the code and the model could not be evolved 
past X$_{\rm C}\simeq 0.18$. 
The solid line shows the same calculation assuming an enhancement in the efficiency of convective 
energy transport (MLT++). 
In this case no envelope inflation occurs and the code can evolve the star to the end of the main sequence.
}
\label{fig:radius}
\end{center}
\end{figure}

In stellar evolution calculations crude approximations are made in order to deal with the 
energy transport in these complex layers which largely affect the final radii of the stars  \citep{2013ApJS..208....4P}.  
For example, in the outer layers of the envelope the temperature decreases to few $\times 10^5$K 
and iron recombines, leading to a large opacity \citep[see e.g.,][]{Cantiello:2009}. 
As a result, in these layers the stellar luminosity 
is much larger than the Eddington luminosity, which might seem to imply that convection will dominate 
energy transport. However, due to the very low density in the envelope, convection is very inefficient 
(superadiabatic) and cannot transport all of the energy, leaving some of it ``trapped" in the envelope. 
A possible (but not unique) solution is that these layers expand, which changes the opacity profile 
and allows the trapped energy to escape. This leads to an inflation of the stellar radius and the 
formation of a density inversion (e.g., \citealt{2006A&A...450..219P,2012A&A...538A..40G}). We note 
that in some codes this inflation does not occur because, for numerical reasons, convection is assumed 
to be able to transport the entire energy flux \citep[see e.g.,][]{1987A&A...173..247M,2013ApJS..208....4P}.
This effect is shown in Fig.~\ref{fig:radius} where the photospheric radius evolution during the main sequence of a 500$\msun$  at $Z=0.002$ is calculated using two different assumptions
for the efficiency of convective energy transport. These models have been calculated using MESA with the usual mixing length theory (MLT) or assuming an 
enhancement of the convective energy transport \citep[MLT++, see Sec.~7.2 in][]{2013ApJS..208....4P}. 
These models are non-rotating and assume the mass loss recipe of \citet{Glebbeek:2009}. 
The result shown by the solid line is in good agreement with a similar calculation using the {\tt GENEVA} code \citep{2013MNRAS.433.1114Y}.

It is estimated \citep{2006A&A...450..219P,2012A&A...538A..40G} that the mass contained in the shells that 
would be affected by this inflation 
is rather small: $10^{-9}\msun$ for Wolf-Rayet stars and $10^{-2}\msun$ for luminous blue variable stars. 
However, the inflation can extend the radius of a star by a factor of a few. This in turn may lead to the 
onset of a CE phase for very massive binaries on relatively close orbits. Once the CE phase is initiated 
we require a significant envelope mass (in excess of $10\msun$) to efficiently decrease the orbital 
separation and form a close BH-BH system. For example, if we start with a massive binary ($400\msun\,+\,200\msun$) 
in an orbit with $a=340\rsun$, the orbital separation will decrease to $290, 120,
14\rsun$ for envelope masses of 
$1, 10, 100\msun$, respectively. Here we have applied the most effective orbital contraction (by assuming energy balance) 
with fully efficient transfer of orbital energy to the envelope ($\alpha_{\rm CE}=1.0$) and with large 
envelope binding energy ($\lambda=0.1$).

Therefore it appears that very massive stars at low metallicity may or may not have massive H-rich envelopes 
(due to the uncertainties in the treatment of internal mixing and wind mass loss rates) and that these envelopes 
may or may not expand significantly after the main sequence (due to the uncertainties in the modeling of 
radiation-dominated stellar envelopes).  To assess the influence of these uncertainties on our predictions we 
consider two extremes in Section~3. For the first we employ evolutionary models for very massive stars that show 
both significant expansion and large envelope mass beyond the main sequence. For the second we assume that very 
massive stars do not expand at all and therefore have no interactions in isolated (e.g., field stellar populations) 
binary evolution.  

\subsection{Black-hole formation}

For stars above $100M_\odot$ the mass of the star at the point of core collapse generally determines the remnant 
mass. If the Carbon/Oxygen (CO) core mass is above $\sim7.6\,M_\odot$ \citep[see, e.g., Section~2.3 
of][]{2010ApJ...714.1217B}, it is likely that the core will collapse to a BH without producing a supernova 
explosion, so the mass of the compact remnant is simply the mass of the star at collapse.  However, if the core 
has sufficient angular momentum, a disk can form around the newly formed BH with the potential to produce 
a gamma-ray burst  \citep{1993ApJ...405..273W,1999ApJ...524..262M}. In a series of $250\,M_\odot$ star simulations, 
assuming efficient internal angular momentum transport from magnetic torques, \citet{2012ApJ...752...32W} found 
that even without mass loss (and the resulting angular momentum loss), only the outer layers (containing $\sim16\,M_\odot$) 
of  these stars have sufficient angular momentum to produce a disk and drive an outflow.  Any such outflow will have 
a small effect ($<5$--$10$\%) on the final BH mass.  Overall our uncertainty about the mass loss from stellar 
winds dominates the uncertainties in mass estimates of the star at collapse and therefore of its compact remnant. 

Very massive stars can produce massive helium cores (above $45\msun$) that, after helium burning, can undergo an 
instability where pair-production reduces the pressure in the core, allowing it to collapse. Compressional heating in 
the collapse causes the CO and Si cores to reach much higher temperatures than occurs in normal core burning, and the 
resulting burning phase can be so explosive that, in the most extreme cases, there is no remnant.  These are the 
hypothesized pair-instability supernova 
\citep{1967PhRvL..18..379B,1984ApJ...280..825B,1984ApJ...277..445C,1985A&A...149..413G,2001ApJ...550..372F,2013ApJ...776..129C}.  
For a sufficiently massive core, the center of the star is so hot that the photodisintegration instability is 
encountered before explosive burning reverses the shock, accelerating the collapse \citep{1984ApJ...280..825B,2001ApJ...550..372F}. 
These stars collapse to form large BHs and the mass at which this photodisintegration instability prevents 
explosions marks the upper limit on the pair-instability progenitor mass. Whether or not the pair instability can 
disrupt a star depends upon the conditions in the core, in particular the helium core mass, which as discussed 
above depends on the details of stellar mixing.  The lower and upper initial star mass limits for pair-instability 
supernovae are typically thought to be roughly $150\msun$ and $300\msun$ \citep{2001ApJ...550..372F}, respectively, 
but these limits are sensitive to the aforementioned uncertainties  on the size of the helium core. To avoid 
confusion with the initial mass range dependence on metallicity, the above limits are typically translated to final 
CO core masses of $60, 130\msun$ (e.g., \citealt{2013MNRAS.433.1114Y}) for the pair-instability mechanism 
to operate. 

Natal kicks during supernovae are another source of uncertainty in this modeling.  It may happen that during core collapse a natal kick will modify the orbit in such a way as to increase the 
eccentricity and/or decrease the orbital separation.  Potentially, a close double BH binary may form. 
However, it is very unlikely that such a favorable kick is encountered, so only a very small fraction of 
binaries may be affected by this process. Nonetheless, such effects could lead to the formation of massive BH-BH 
binaries in some cases. The tremendous reach of advanced ground-based instruments means that such coalescences need 
only happen a few percent of the time to lead to many detections per year. 

Supernova natal kicks may play an additional role in these massive stars. Without supernova explosions, kick 
mechanisms relying on asymmetric mass ejecta (e.g., \citealt{1992ApJ...395..642H,2003PhRvL..90x1101B,2003ApJ...584..971B}) 
will not work.  In the absence of kicks, we expect the spin axes to be aligned for the resulting binary BH systems. 
This is because binary stars {\it (i)} may have been formed with rotation aligned with the orbital axis 
(although see \citep{Albrecht2014} for examples of non-aligned binaries), {\it (ii)} may have been subject to 
strong tidal interactions if on relatively close orbits (although see \citep{Claret2007} for a discussion of very inefficient radiative damping and tidal torquing for massive stars) and {\it (iii)} mass transfer and/or common envelope episodes may have aligned component rotation with the orbital angular momentum for close interacting 
binaries (although see \citep{Sepinsky2010} for examples when mass transfer does not necessarily lead to effective tidal torquing and circularization/synchronization).
If this is the case, then the alignment of spin axes could be 
used to distinguish a binary origin of massive BH-BH mergers from a dynamical origin.  However, mechanisms invoking 
asymmetries in the neutrino emission (e.g., \citealt{1996PhRvL..77.4872K,1997PhLB..396..197K,2005ApJ...632..531S,2006ApJS..163..335F}) 
are able to drive kicks as long as electron neutrinos are trapped in the collapsed core.  These mechanisms require 
large magnetic fields ($\gtrsim 10^{15}$G). For massive stars with zero-age main sequence masses above 1,000\,M$_\odot$, the electron 
neutrino trapping region is small, but below this mass (which encompasses all the systems of interest in this paper), 
the trapping region can be large \citep{2011AN....332..408F}. It is therefore possible that a neutrino driven kick 
mechanism may work. Such kicks would alter our population studies  and produce non-aligned spin axes and would reduce 
close BH-BH formation rates. Without a better quantitative understanding of these kicks, however, it is difficult to 
assess their effects.

\section{Estimates of BH-BH merger/formation rates} 

In this section we give the results from two models that give a reasonable span 
of possible rates.  In Model~1 we assume that very massive stars expand 
significantly and have massive H-rich envelopes beyond the main sequence. In 
this approach we use population synthesis to estimate the merger rates of massive 
close BH-BH binaries. In Model~2 we assume that very massive stars do not expand 
significantly beyond the main sequence and we use simple order-of-magnitude 
estimates to assess the formation rates of wide massive BH-BH binaries. In both 
models we use an energy balance approach for the CE phase.  We also limit the 
range of pair-instability supernovae to stars that form final CO core masses of 
$60$--$130\msun$ and assume that in this range no BHs are formed.  Outside this 
range we assume that the entire star collapses to form a BH (minus the $10\%$ of 
the mass that we assume leaves in the form of neutrino emission).  Therefore, we 
neglect any potential mass loss from collapsar outbursts. We also assume that these 
massive stars do not impart natal kicks during the formation of their BHs.  
In Section~4 we will explore dynamical processes to show that these may shorten the 
merger time of these wide BH-BH binaries below a Hubble time and make them 
potentially important for advanced GW detectors.  

\subsection{Model 1: expanding VMS}

\subsubsection{Initial Conditions}

Recent evidence suggests that very massive stars exist above $1$--$10$\% 
solar metallicity (where $Z_\odot=0.014$), but little is 
known yet about their initial conditions in binaries, e.g., their initial mass function, 
binary fractions, and orbital separations. We assume that the stellar initial mass 
function for the primary mass $M_{\rm zams,a}$ has the form $dN/dM\propto M^{-\alpha}$ 
with $\alpha=1.3$ for the range $0.08\,M_\odot$--$0.5\,M_\odot$, $\alpha=2.35$ for 
$0.5\,M_\odot$--$1\,M_\odot$, and $\alpha=2.7$ for $M>1\,M_\odot$ 
\citep{1993MNRAS.262..545K,2011MNRAS.412..979W}.  
We only consider here very massive stars with $150<M_{\rm zams,a}<1000\msun$.
We also assume a companion mass $M_{\rm zams,b}\leq M_{\rm zams,a}$ drawn from a 
distribution uniform between $0$ and $1$ in the mass ratio $q\equiv M_{\rm zams,b}/M_{\rm zams,a}$ 
\citep[e.g.,][]{2012Sci...337..444S}, a uniform-in-the-log distribution of orbital separations
and a thermal-equilibrium distribution of eccentricities ($P(e)= 2e$ in range
$e=0-1$; \cite{1975MNRAS.173..729H}; \cite{Duquennoy1991}). 
If very massive stars expanded as much as $M_{\rm zams} < 150 M_\odot$ stars, these distributions 
would ensure that $\sim 50\%$ of binaries would experience Roche lobe overflow during their lifetime, 
leading to a potential common envelope and subsequent orbital contraction \citep{2013A&A...550A.107S}; 
however, the lack of significant expansion of very massive stars may limit such interactions.
We only evolve binaries for which $M_{\rm zams,b}>150\msun$ as in this study we are focused on very 
massive star evolution and the formation of massive BH-BH binaries.  We assume a binary 
fraction of $50\%$ (i.e., $2/3$ of stars are in binaries). We also assume that
$50\%$ of the stars in the local universe (within the reach of advanced GW 
instruments) have solar metallicity and $50\%$ of the local stars have $Z=0.002$ 
($\sim 10\% \zsun$). 
 
Some of these assumptions have linear effects on the predicted rate. For example, 
changing the number of massive stars by a factor of two will change the number of 
mergers by a factor of two, and changing the close binary fraction by a factor of 
two changes the number of mergers by roughly a factor of two. In contrast, our 
uncertainties about stellar evolution and binary interactions have non-linear 
effects, and will dominate the errors in our rate estimates for the foreseeable 
future.

\subsubsection{Population Synthesis Calculations}

Binary evolution calculations are treated as described in full detail by
\cite{2008ApJS..174..223B} with several recent updates on stellar wind mass
loss rates, compact object mass distribution and CE handling as outlined by 
\cite{2012ApJ...759...52D}. In particular, we employ wind mass loss rates
from \cite{2001A&A...369..574V}. We use energy balance for CE evolution 
that employs fully efficient transfer of orbital energy into the CE (i.e.,
$\alpha=1$) with physical calibration of donor binding energy \citep{Xu2010}. 
A criterion for the development of CE is based on donor expansion/contraction 
in response to mass loss and the related response of donor Roche lobe
due to mass transfer and orbital angular momentum losses. This approximately
translates into CE development for binaries with donors significantly (factor
of $\gtrsim 2$) more massive than their companions or with donors with deep 
convective envelopes (e.g., red giants).
We apply the rapid supernova explosion model, which reproduces the mass gap
between neutron stars and BHs, to obtain compact remnant masses 
\citep{2012ApJ...757...91B}.

The community currently lacks fully accepted models of very massive stars at high 
metallicity.  In lieu of such models, we extend the stellar evolutionary formulae of 
\citet{2000MNRAS.315..543H} to calculate the evolution and fate of stars up to a ZAMS 
mass of $M_{\rm zams}=1,000 \msun$ using the population synthesis code \STRACK 
\citep{2008ApJS..174..223B}. We apply a direct extrapolation with no high-mass 
specific modifications to the original \citet{2000MNRAS.315..543H} models. We combine 
these formulae with the wind mass loss rates compiled and calibrated by 
\citet{2010ApJ...714.1217B}. This scheme was originally developed for stars with 
$M_{\rm zams} \lesssim 100$--$150 \msun$. These extrapolated VMS models expand 
significantly and form massive H-rich envelopes beyond the main sequence, as expected 
for stars with $M_{\rm zams} \lesssim 100$--$150 \msun$. 

\begin{figure}
\begin{center}
\includegraphics*[width=0.7\textwidth]{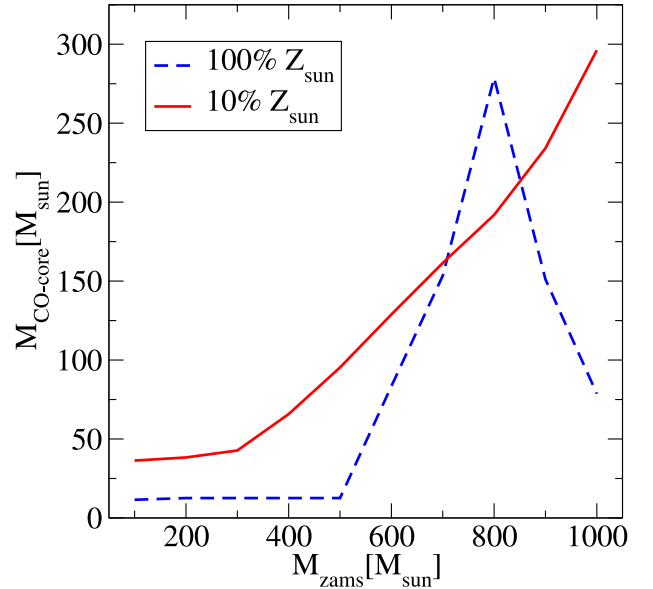}
\caption{Final CO core mass (just prior to core collapse and BH formation) for our 
evolutionary models of very massive stars. Here the stars are assumed to be either 
single or non-interacting binary components, so this figure is only illustrative and is 
not directly relevant for binary evolution leading to the formation of close BH-BH 
systems.
}
\label{fig:cocoremass}
\end{center}
\end{figure}

In our solar metallicity models the CO core mass is approximately $12 \msun$ for a 
broad range of initial stellar masses, $M_{\rm zams}=100$--$500 \msun$ 
(Fig.~\ref{fig:cocoremass}).  At higher ZAMS masses the CO core mass rises sharply to 
a maximum of $\sim 280 \msun$ for $M_{\rm zams}=800\msun$, then decreases for higher 
initial star masses. At subsolar metallicity, the CO core mass is roughly $40 \msun$ 
for $M_{\rm zams}=100$--$300\msun$, above which the core mass rises monotonically to 
a maximum of $300 \msun$ for $M_{\rm zams}=1,000\msun$.  As we indicated in Section~2, 
the BH mass approximately traces the CO core mass, and so depends on the initial 
stellar mass in the same way that the CO core mass does, for both solar and subsolar 
metallicity. However, whereas the maximum CO core mass is roughly the same for solar 
and subsolar models, the maximum BH mass is much greater for subsolar metallicity stars. 
This is because at subsolar metallicity the highest mass stars retain their H-rich 
envelope, whereas at solar metallicity winds can efficiently remove the entire envelope.

Our calculation of CO core mass is based on the combination of extrapolated stellar 
models presented by \citet{2000MNRAS.315..543H} and the wind mass loss rates collected 
and calibrated by \citet{2010ApJ...714.1217B}. Within this framework the initial flatness 
of the dependence of CO core mass on ZAMS mass is the result of increasing strength of 
wind mass loss with ZAMS mass. Stars below $M_{\rm zams}=500\msun$ and $M_{\rm zams}=300\msun$ 
for solar and subsolar metallicity, respectively, are stripped of their H-rich envelopes and 
they become WR stars. This further increases mass loss. Low metallicity stars above 
$M_{\rm zams}=300\msun$ have such massive H-rich envelopes that they never become WR stars, 
and their pre-supernova mass (and therefore CO core mass) increases monotonically with ZAMS 
mass. This happens also for solar metallicity stars above $M_{\rm zams}=500\msun$. However, 
at some point, high metallicity stars are subject to wind mass loss strong enough to remove 
even very massive H-rich envelope and they become massive WR stars. Strong WR-type wind mass 
loss leads to the efficient decrease of star mass. For stars above $M_{\rm zams}=800\msun$ 
the final CO core mass decreases with initial ZAMS mass.

These results may be compared to the detailed stellar evolutionary models of massive 
stars published by \citet{2012IAUS..279..431Y}. They found that at solar metallicity 
the final CO cores have masses $15$--$20 \msun$ (assuming no rotation) or around 
$25$--$35 \msun$ (with significant rotation), for $M_{\rm zams}=100$--$500 \msun$. 
Our relation for the same regime is also flat but results in lower CO core masses 
($12 \msun$). Their lowest metallicity model, which has a metallicity of $\sim 10\%$ 
solar ($Z=0.002$; appropriate for the Small Magellanic Cloud), shows a monotonic 
increase of final CO core mass from $90\msun$ to $140\msun$ in the mass range 
$M_{\rm zams}=150$--$300 \msun$.  Our relation is flatter in that regime and results 
in lower CO core masses ($40$--$50 \msun$). In both solar and subsolar metallicity 
evolution our evolutionary models result in lower CO core masses than are predicted 
by \citet{2012IAUS..279..431Y}. Therefore, our estimate of BH mass is conservatively 
low compared with the published detailed evolutionary models (see Fig.~\ref{fig:cocoremass}). 
We also stress that for the relatively high metallicities, $Z>0.4~Z_\odot$, considered 
by \citet{2012IAUS..279..431Y}, winds drive away the entire H-rich envelope for high-mass 
stars. This is consistent with our findings, but may not be true for lower metallicity stars.

\begin{deluxetable}{lcccl}
\tablewidth{0pc}
\tablecolumns{5}
\tablecaption{Very Massive Star Model Properties\tablenotemark{a}}
\tablehead{%
\colhead{Phase} & 
\colhead{Mass} &
\colhead{Radius} &
\colhead{ $M_{\rm CO}$} &
\colhead{ Comment} \\
 \colhead{} &
\colhead{M$_\odot$} &
\colhead{R$_\odot$} &
\colhead{M$_\odot$} &
 \colhead{} 
}
\startdata
start MS         & 500  &  50    &  --  & $Z=0.014$   \\   
end MS           &  48  &   5    &  --  &               \\
end He-burning   &  26  &   1    &  20  & BH formation  \\
                 &      &        &      &               \\

start MS         & 500  &  50    &  --  & $Z=0.006$   \\   
mid MS           &      & 100    &  --  &                   \\  
end MS           & 102  &   7    &  --  &                   \\
end He-burning   &  75  &   2    &  65  & Pair inst. SN     \\
                 &      &        &      &                   \\

start MS         & 500  &  30    &  --  & $Z=0.002$  \\   
mid MS           &      &  50    &  --  &                   \\  
end MS           &      &  10    &  --  &                   \\
post MS          &      & 100    &  --  &                   \\ 
end He-burning   &      & $<65$  &  150 & BH formation      \\

\enddata
\label{sn}
\tablenotetext{a}{These results are taken (or extrapolated) from the rotating \citet{2013MNRAS.433.1114Y} 
\vspace*{0.4cm}models.}
\label{vms}
\end{deluxetable}

We can also compare our results to the additional models in \citet{2013MNRAS.433.1114Y}.  
Here we will focus on the highest mass models in these calculations, with an 
initial mass of $M_{\rm zams}=500\msun$; somewhat lower-mass stars (e.g., 
$M_{\rm zams}=300\,M_\odot$) evolve in qualitatively similar ways. We will use 
only their rotating models, as these are more physical than non-rotating models. 
Table~\ref{vms} shows the results from the calculations in \citet{2013MNRAS.433.1114Y}.  
For the solar-metallicity $500\msun$ star, the mass of the star is $26\msun$ at the end 
of core helium burning, dropping to $20\msun$ when the CO core forms. Such a star will 
most likely collapse to a BH with mass similar to the mass of its CO core mass. At about 
$40\%$ solar metallicity ($Z=0.006$; typical of the Large Magellanic Cloud), the mass of 
the star at the end of the main sequence is $102\msun$ and  at the end of core helium 
burning is $75\msun$. The star forms a $65\msun$ CO core and at this mass it is likely to 
undergo a pair instability supernova that completely disrupts the star and does not leave 
behind a compact object remnant. For $10\%$ of solar metallicity, extrapolating the results 
from \citet{2013MNRAS.433.1114Y} suggests that a $M_{\rm zams}=500\msun$ star ultimately 
produces a CO core mass of $150$--$170 \msun$. The most likely fate of such a star is core 
collapse and the formation of a BH with a mass of about $150\msun$. For comparison, a 
$300\msun$ model at this metallicity finishes the main sequence as a $176\msun$ star, and 
ends core helium burning as a $150\msun$ star that forms a CO core with a mass of $135\msun$. 
This CO core will likely form a BH of similar mass.

\subsubsection{Population Synthesis Results}
\label{sec:popsyn}

We find that only low metallicity environments/host galaxies could contribute significantly to 
the formation of close BH-BH systems from very massive stars. This is qualitatively the same 
result as obtained for lower mass stars \citep{Belczynski2010a}. Most of the close very 
massive BH-BH systems ($75\%$) merge with short delay times ($t_{\rm delay} < 1$ Gyr). 

For subsolar metallicity there are two separate populations of BH-BH systems. Most of the 
systems ($93\%$) form with total mass $M_{\rm tot} = 50$--$100 \msun$ and average mass ratio of 
$q=0.9$, while there is a smaller population ($7\%$) of BH-BH systems with 
$M_{\rm tot} = 100$--$300 \msun$ and average mass ratio of $q=0.8$ for subsolar metallicity. The 
less massive systems originate from stars with $M_{\rm zams} \approx 150$--$500 \msun$, whereas 
the more massive ones evolve from stars with $M_{\rm zams} \approx 500$--$1000\msun$.  
For solar metallicity, there are virtually no BH-BH systems with total mass 
$M_{\rm tot} > 40 \msun$.  

\begin{figure}
\begin{center}
\includegraphics*[width=0.47\textwidth]{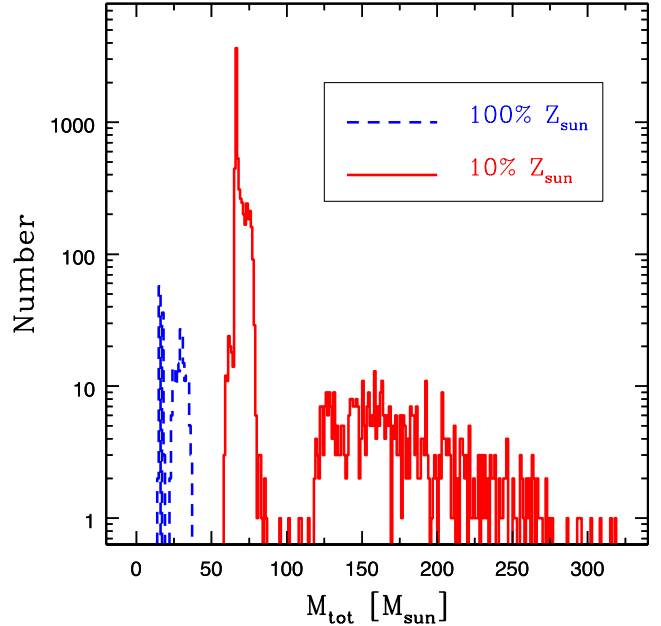}
\caption{Total mass of close (merging within a Hubble time) BH-BH binaries produced in 
our Model $1$ with population synthesis calculations. Note that the low metallicity environment
dominates BH-BH formation. Also note two distinctive BH-BH subpopulations at low metallicity: 
one with $M_{\rm tot}<100\msun$ and one with $M_{\rm tot}>100\msun$.}
\label{fig:mtot}
\end{center}
\end{figure}

\begin{figure}
\begin{center}
\includegraphics*[width=0.47\textwidth]{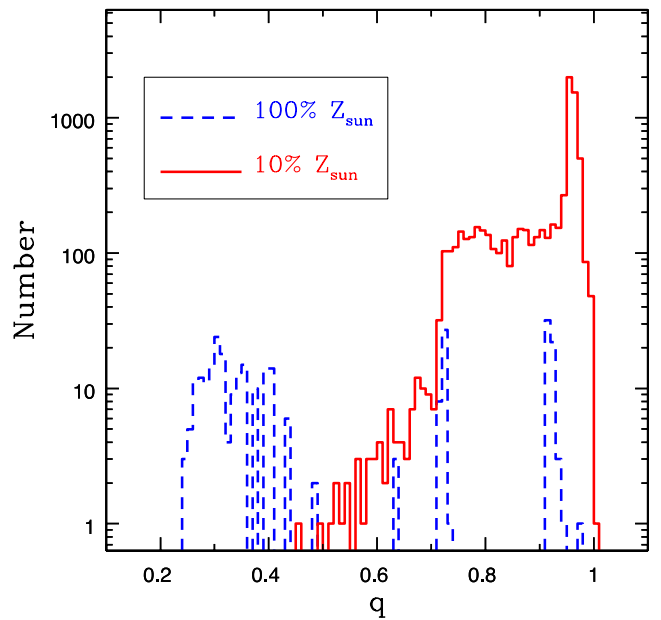}
\caption{Close BH-BH binary mass ratio (less massive to more massive) for our Model $1$. 
Note that mass ratios are rather high and are typically in range $q=0.7-1$ for the
dominant low metallicity BH-BH population.}
\label{fig:q}
\end{center}
\end{figure}

\begin{figure}
\begin{center}
\includegraphics*[width=0.47\textwidth]{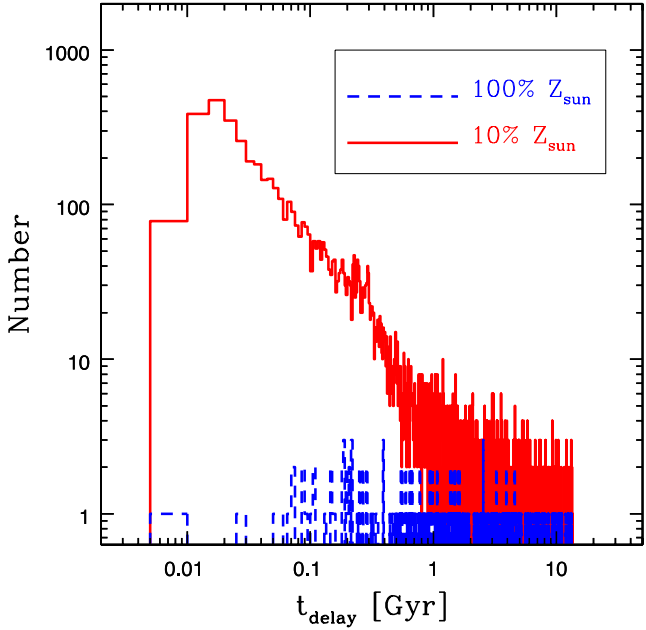}
\caption{Delay time distribution of close BH-BH binaries for our Model $1$. 
Note that the distribution for the dominant low metallicity BH-BH population
falls off approximately as $t_{\rm delay}^{-1}$. The delay time is defined
as time elapsed from star formation (ZAMS) to the final BH-BH coalescence.}
\label{fig:tdel}
\end{center}
\end{figure}

We consider only close BH-BH binaries with total mass $M_{\rm tot} > 100 \msun$, as these
binaries provide a clear signature of VMS existence and evolution. The lower mass systems 
may have originated from regular stars (i.e., $M_{\rm zams} < 150 \msun$). These massive 
BH-BH systems would have aligned spins (as no supernova explosion and hence no natal kick 
is expected for such massive stars; \citealt{2012ApJ...749...91F}). If core collapse for 
such massive stars leads to a stalled shock rather than mass loss in a supernova, the BHs 
that are formed will include the entire angular momentum of the pre-collapse progenitor.  
This is likely to lead to rapidly spinning BHs.
Our major results discussed above are presented in Figure~\ref{fig:mtot}, 
~\ref{fig:q} and ~\ref{fig:tdel}.

With all of the previously discussed uncertainties, it is very difficult to produce reliable 
quantitative results for BH binaries. However, within the framework of these 
extrapolated normal stellar models we find that the common envelope phase is almost always 
initiated by a star when it expands immediately after the main sequence, i.e., during 
Hertzsprung gap evolution. If we assume the binary can survive this common envelope phase 
and that pair instability supernovae do not occur for Population I stars ($Z>0.1 \zsun$, which 
encompasses all our models), we find that BH-BH merger rates are $0.5 \times 10^{-6}$ 
yr$^{-1}$ per Milky Way equivalent galaxy (MWEG) for BH-BH systems with total masses 
larger than 100 $M_\odot$. This corresponds to a merger rate density of $5 \times 10^{-9}$ 
yr$^{-1}$ Mpc$^{-3}$. This is our most optimistic estimate. If we allow for pair 
instability supernovae as discussed in Section~2 then the rates drop by two orders of 
magnitude and we find that the BH-BH merger rate density is $5 \times 10^{-11}$ yr$^{-1}$ 
Mpc$^{-3}$. This is probably our most realistic model. 
If we increase our CO core masses by a factor of two (consistent with \cite{2013MNRAS.433.1114Y},  
who find CO cores $\sim 1.5-3$ times more massive than ours), the rate increases to  
$3\times 10^{-9}{\rm yr}^{-1}~{\rm Mpc}^{-3}$. This is because our CO core masses for  
close BH-BH systems with $M_{\rm tot}>100~M_\odot$ are $M_{\rm CO}=50-100~M_\odot$, which 
is in the pair instability supernova range $M_{\rm CO,pair}=60-130~M_\odot$. Doubling our  
CO core masses therefore moves most of them above the pair instability range and thus 
allows many additional systems to evolve to BH-BH binaries. 
Finally, if we do not allow for CE survival with Hertzsprung gap donors the merger rates 
drop to zero.

We note that all of our findings in this section are a direct result of our {\it assumption} 
that very massive stars of high metallicity behave qualitatively like normal massive stars.  
If they do, then the large volume in which massive BH coalescence could be seen with 
advanced LIGO/Virgo suggests that the observed rate could be many per year. The observation of 
even one of these systems would refute the arguments against radial expansion of very massive 
stars \citep{2013MNRAS.433.1114Y}, and would show that these binaries can survive the CE 
phase with Hertzsprung gap donors. These constraints would provide insight into stellar 
evolution models. If the rate is high, we could even place limits on the formation conditions 
of these massive binaries (e.g., the binary mass ratio and initial mass function).

\subsection{Model 2: non-expanding VMS}
\label{sec:OoM}

In this model we assume that there is no significant expansion of VMSs. This means that both 
binary components evolve as single stars, and in the end either form BHs or are disrupted by 
pair-instability SNe. The initial orbital separation only expands due to the wind mass loss. 
There is no CE phase that could decrease orbital separation as both binary components are within 
their respective Roche lobes. Since we assume no natal kicks at the formation of the BHs, all 
binaries with components that are not subject to pair-instability SNe should form wide BH-BH 
binaries (i.e., with merger times much longer than the Hubble time). However, as we discuss in
Section~4, dynamical processes may be able to tighten such binaries so that they do merge
within a Hubble time.  Here we present an 
order-of-magnitude estimate of the formation rate of these wide BH-BH systems.  We will limit 
our estimate only to systems with BHs of $100\msun$ or more\footnote{Note that in Model 1 we 
have only considered BH-BH binaries with total mass above $100\msun$, while the Model 1 total 
mass distribution extended to about $300\msun$.}.

We make the same assumptions about the initial mass function as before,
extending it to an upper limit of $10^3\,M_\odot$. The wind-driven mass loss rate for high-mass 
stars is extremely uncertain, as is the dependence of this rate on metallicity.  It is therefore 
unclear what initial mass is required to leave behind a $100\,M_\odot$ BH, but for an 
order-of-magnitude estimate we will assume that $M_{\rm zams}>500\,M_\odot$ is sufficient to form 
a $100\,M_\odot$ BH. For comparison, our extrapolated evolutionary models produce a $100\,M_\odot$ 
BH for $M_{\rm zams}>600\,M_\odot$ and $M_{\rm zams}>400\,M_\odot$ for solar and $10\%$ solar 
metallicity, respectively.  For the initial mass function we have assumed, we find a fraction 
$\approx 2\times 10^{-6}$ of stars will start with masses above $500\,M_\odot$. If, as we assumed 
previously, the binary mass ratio distribution is roughly flat, then $\sim$half of the stars will 
have companions at least half as massive as they are, and thus binary evolution will presumably 
produce comparable-mass BHs.

Let us suppose that only some fraction of these binaries can avoid pair-instability supernova.  
Following our earlier discussion we assume that stars with final CO core mass above $130\msun$ 
avoid this fate and form massive BHs.  Extrapolation of results presented by \citet{2012IAUS..279..431Y} 
(see their Fig. 18) suggests that only stars at subsolar metallicity can avoid pair 
instability SNe and form massive BHs above $100\msun$. It appears that at $Z=0.002$ (SMC) all stars 
with $M_{\rm zams}>500\msun$ form these massive BHs, whereas only a small fraction (if any) of stars 
with $M_{\rm zams}>500\msun$ form massive BHs for $Z=0.006$ (such as for the Large Magellanic Cloud). 
At higher metallicities (e.g., for the Milky Way) no massive BHs are expected. Given that the 
metallicity distribution of stars is poorly constrained, we assume that only $10^{-1}$ can 
avoid pair-instability SNe for  $M_{\rm zams}>500\msun$.

The Galaxy has $\sim 2\times 10^{11}$ stars, so our collection of assumptions implies that 
$\sim 2\times 10^4$ wide very massive BH binaries per Milky Way Equivalent Galaxy (MWEG) will be 
produced by a stellar binary population. The Galaxy is $\sim 10^{10}$~years old, and the comoving 
number density of MWEGs is $\sim 10^{-2}~{\rm Mpc}^{-3}$ (e.g., \citealt{2008ApJ...675.1459K}).  
Thus the formation rate density, averaged over cosmic history, is $\sim 10^{-8}~{\rm Mpc}^{-3}~{\rm yr}^{-1}$. 
In Section~5 we discuss how these rates translate into advanced LIGO detection rates.

We can similarly estimate the formation rate of $\gtrsim
  100\,M_\odot$ BHs with $\sim 10\,M_\odot$ BH companions. These
  intermediate-mass-ratio binaries are of significant interest as
  their inspirals may serve as precise probes of the spacetime around
  BHs
  \citep{2007CQGra..24R.113A,2007PhRvL..99t1102B,2011GReGr..43..485G,2012PhRvD..85f2002R}. If
  we continue to assume that the mass ratio distribution is flat, then
  we would expect $\sim 10\,M_\odot$ BHs to be companions to
  $>100\,M_\odot$ BHs in $\sim 10$\% of systems, leading to a formation rate of $\sim 10^{-9}\,{\rm Mpc}^{-3}\,{\rm yr}^{-1}$.

\section{Dynamical processes affecting massive binaries}
\label{sec:dynamics}

As we have discussed, current stellar evolutionary theory suggests that although massive main 
sequence binaries have a good chance of evolving into massive BH binaries, those binaries 
seem unlikely to be close enough that, on their own, they will coalesce within a Hubble time.  
However, many or most massive stars are born in stellar associations with initial number densities 
of hundreds of stars or more per cubic parsec.  Hence after the formation of a double BH 
system there are additional interactions that can harden the binary.  In addition, there is evidence 
that massive stars are, in a few to $>10$\% of cases, in triple or higher-order systems (e.g., 
\citealt{2012ApJ...747...41K,2012ApJ...751....4K,2013arXiv1303.3028D}).  Such systems can potentially 
undergo Kozai resonance cycles \citep{1962AJ.....67..591K,1962P&SS....9..719L} that increase the 
eccentricity of the massive binary sufficiently so that orbital energy and angular momentum loss through the 
emission of GWs could cause rapid coalescence.  In this section we examine these effects 
and show that a significant fraction of initially wide very massive binaries may merge within a Hubble 
time, and possibly much sooner.

\subsection{Binary-single interactions}

Suppose that our massive binary is born into a cluster with a central mass density of 
$\rho \equiv 10^3\rho_3\,M_\odot\,{\rm pc}^{-3}$; note that R136 has a central density of at least 
$1.5\times 10^4\,M_\odot\,{\rm pc}^{-3}$ \citep{2013A&A...552A..94S}, and that other young clusters 
such as the Arches cluster can be even denser \citep{2001ApJ...551L.143L}.  If our massive binary 
has a total mass $M$ and a semimajor axis $a$, and if we assume that the single objects that 
encounter the binary have much lower mass and need to get to within a distance $\sim a$ of the 
center of mass of the binary to affect the binary semimajor axis and eccentricity, then for a 
relative speed $v_\infty$ at infinity the cross section for the interaction is
\begin{equation}
\Sigma=\pi a^2\left({2 GM\over{av_\infty^2}}+1\right)\; .
\end{equation}
Stellar associations typically have velocity dispersions of a few km~s$^{-1}$, so the first term in 
the parentheses dominates as long as $a$ is less than $\sim 10^3$~AU.  For the same reason, 
interactions of interloping stars with the binary will tend to harden the binary and change its 
eccentricity $e$, causing it to sample a ``thermal" distribution $P(e)de=2ede$ (see 
\citealt{1975MNRAS.173..729H} for a pioneering study).  Interaction with interlopers with a total 
mass approximately $2\pi/22$ times the binary's mass will reduce the semimajor axis by a factor of 
$\sim 3$ \citep{1996NewA....1...35Q}.  The time needed to interact with this much mass is
\begin{equation}
T={M\over{\rho\Sigma v_\infty}}\approx 6\times 10^8\, {\rm yr}\  
\rho_3^{-1}\left(\frac{a}{10~{\rm AU}}\right)^{-1}\left(\frac{v_\infty}{3~{\rm km~s}^{-1}}\right).
\end{equation}
Note that because $\Sigma\propto v_\infty^{-2}$, a smaller velocity dispersion implies a more rapid 
hardening of the binary.  Note also that $M$ does not appear, because the greater number of 
interactions required is offset by the larger gravitationally-focused cross section.  If the 
interloping objects are sufficiently massive then significant kicks could be delivered to the binary 
in the binary-single interactions, but if the binary is much more massive than typical stars (which 
we assume to be the case) then the binary should be able to remain within the stellar cluster. We 
also note that because massive objects tend to swap into binaries, initially solitary massive BHs 
have a good chance to become members of binaries in a short time (again, see \citealt{1975MNRAS.173..729H} 
for an early discussion of this process).

If the binary orbit were forced to remain circular then its GW inspiral time is 
\citep{1964PhRv..136.1224P}
\begin{equation}
\begin{array}{rl}
T_{\rm GW}(e=0)&=5a^4c^5/(256G^3\eta M^3)\\
&\approx
1.6\times 10^{15}\,{\rm yr}\left(\eta\over{0.25}\right)^{-1}\left(M\over{200\,M_\odot}\right)^{-3}
\left(a\over{10\,{\rm AU}}\right)^4 \\
\end{array}
\end{equation}
where $\eta=m_1m_2/M^2$ is the symmetric mass ratio for objects of masses $m_1$ and $m_2$ and total mass 
$M$; we have normalized $\eta$ to $0.25$, which is its maximum value and which occurs for equal masses.  
For a binary of eccentricity $e$ that is reasonably near unity, this time becomes \citep{1964PhRv..136.1224P}
\begin{equation}\label{TGWe}
T_{\rm GW}(e\approx 1) \approx (768/425)(1-e^2)^{7/2}T_{\rm GW}(e=0)\; .
\end{equation}
Thus for a binary of two $100\,M_\odot$ BHs in a 1\,AU orbit, an eccentricity of 
$\approx 0.895$ [$0.88$ if the full expression from \citep{1964PhRv..136.1224P} is used rather than the approximation (\ref{TGWe})] is enough to drop the merger time to 1\,Gyr.  If the BH binary has $a=10$\,AU, the 
needed eccentricity for a 1\,Gyr merger time is $\approx 0.992$.  

Binary-single interactions will thus drive up the eccentricity until the the inspiral time becomes
short and the binary will 
merge (see \citealt{2006ApJ...640..156G} and \citealt{Mandel:2008} for similar arguments in the case of massive 
BH binaries and intermediate-mass-ratio inspirals, respectively,   
in globular clusters).  As discussed above, the eccentricity distribution ($P(e)=2e$) is sampled roughly 
uniformly in each interaction in the limit of three objects of comparable
mass. In the limit of a 
massive binary and very low-mass interlopers, \citet{1996NewA....1...35Q} showed that three-body interactions 
increase the eccentricity of the binary steadily but slowly.  For example, from Quinlan's formulae, a 
$100\msun$--$100\msun$ binary with $a=10$\,AU in a cluster of velocity dispersion $\sigma=3$\,km\,s$^{-1}$ will 
increase its eccentricity from 0.7 to 0.99 in the time needed to reduce $a$ by 2.4 e-foldings ($\sim 1$\,Gyr 
for our parameters), and from 0.7 to 0.999 in the time needed to reduce $a$ by 3.4 e-foldings.  Eccentricity 
fluctuations due to finite-mass interlopers will likely decrease the time to a given eccentricity.  The 
details depend on the initial distribution of semimajor axes and eccentricities, and on the mass density of 
the cluster and the cluster's longevity, but it is plausible that in a significant fraction of cases where 
massive BH binaries form from stellar evolution, they are induced to merge by repeated dynamical 
encounters.  The tilting of orbits during the dynamical encounters means that the BH spins will 
typically not be aligned with the orbit during merger.

\subsection{Triple systems and the Kozai mechanism}

Another way that initially widely separated massive BH binaries might coalesce involves triple 
systems.  Observed massive stars have a probability, possibly greater than 10\%, of being in triple or 
higher-order systems (e.g., \citealt{2012ApJ...747...41K,2012ApJ...751....4K, 2013arXiv1303.3028D}).  If 
the relative inclination of the inner binary and the tertiary is in the appropriate range, then over many orbits 
of both the binary and the tertiary the mutual inclination will oscillate between two limits in such a way 
that the eccentricity also oscillates \citep{1962AJ.....67..591K,1962P&SS....9..719L,1976CeMec..13..471L}. 
In the standard Kozai approximation, in which the tertiary has a fixed and effectively infinite angular 
momentum, the maximum eccentricity reached during a cycle is $(1-(5/3)\cos^2 i)^{1/2}$, where $i$ is the 
initial relative inclination.  For example, if $i=70^\circ$, the maximum eccentricity is 0.9. If, as is 
likely in our case, the binary has as much or more angular momentum as the tertiary, then the critical 
mutual inclination deviates from $90^\circ$ (see, e.g., \citealt{1976CeMec..13..471L,2002ApJ...576..894M}), 
so the probability of achieving a high eccentricity assuming random inclinations will be somewhat smaller 
than it would be for a critical inclination of $90^\circ$. 

The time for a single Kozai cycle (from minimum to maximum eccentricity and back) is of order 
$\sim (M/m)(b/a)^3$ times the orbital time of the binary, where $M$ is the mass of the binary, $m$ is the 
mass of the tertiary, $b$ is the semiminor axis of the tertiary, and $a$ is the semimajor axis of the 
binary (e.g., \citealt{1976CeMec..13..471L}).  For $a=10$\,AU, $b/a=10$, $M=200\,M_\odot$, and $M/m=100$ 
(reasonable fiducial values), this time is roughly $2\times 10^5$\,yr, which is much less than the time 
between close interactions with interloping stars.  Thus systems that are favorably inclined for the Kozai 
resonance can complete many cycles without being harassed by interlopers.

General relativistic pericenter precession can limit the maximum eccentricity during a Kozai cycle. However, 
from the expressions in \citet{2002ApJ...576..894M}, this is not a limiting factor in our case.  From their 
equation~(6), the maximum eccentricity that can be attained for a binary of two $100\,M_\odot$ BHs at 
1\,AU orbited by a $10\,M_\odot$ object with a semiminor axis of 10\,AU is 0.99, far above the $e\approx 0.9$ 
needed for merger within 1\,Gyr.  If the binary has a semimajor axis of 10\,AU and the tertiary has a semiminor 
axis of 100\,AU, the maximum eccentricity is 0.9999, compared with the $e\approx 0.99$ for a 1\,Gyr merger.  
Note that such an eccentricity can be achieved by $\sim 10$\% of systems\ for an isotropic distribution of 
initial mutual inclinations.

Thus, depending on the fraction of triple systems and the distribution of the masses, inclinations, and 
separations of the tertiaries, it is possible that a few percent of initially wide very massive star binaries 
(tens of percent of the $\sim 10$\% of systems that are triples) are induced to merge by Kozai cycles.  
Such mergers would not have spins aligned with their orbits, because orbital tilting is produced during 
the cycles.  The studies of \citet{2002ApJ...578..775B} suggest that merger will happen near the point 
of minimum mutual inclination between the tertiary and binary, and if this is at a point where the more 
massive BH in the binary is reasonably aligned with the orbital axis and the binary component mass 
ratio is not more than $\sim 2:1$, additional effects during the inspiral could help to align the axes 
further \citep{2004PhRvD..70l4020S}.

In summary: in addition to the possibility of merger through isolated binary
evolution (which remains viable because of uncertainties in massive binary
evolution),
there are at least two dynamical 
scenarios that might produce mergers of a few percent or more of the massive BH binaries that are 
too wide to merge in a Hubble time through radiation reaction alone.  Coalescences through isolated binary 
evolution are likely to yield binary BHs with nearly aligned spins whose magnitude is close to the 
astrophysical maximum, if supernova kicks for these binaries are minimal.  Coalescences involving dynamical 
processes will probably produce unaligned mergers, but because the BHs will have formed before the 
dynamical interactions the spins may still be close to maximal.  Thus detections will probe a variety of 
stellar evolutionary and dynamical processes. We now discuss how coalescences between such objects could be 
detected.

\section{Detection of BH-BH coalescences}
\label{sec:bhbh_coalescence}

Advanced LIGO \citep{AdvLIGO} 
and Virgo \citep{AdvVirgo} interferometric GW detectors are 
coming online within a few years, and are expected to reach design sensitivity by the end of the decade 
\citep{scenarios}.  In addition to improving overall sensitivity by a factor of $\sim 10$ relative to 
their initial versions \citep{LIGO, Virgo}, these detectors will also extend the sensitivity to lower frequencies, 
down to $\sim 10$ Hz, which will allow GW signatures of massive BH   
binary mergers to be detected \citep[see][for upper limits from searches with initial LIGO and Virgo 
detectors]{IMBHB:S5}.   Matched filtering against known templates describing expected signals is the optimal 
technique for extracting weak GW signals  from noisy data.  Although a network of multiple 
detectors is needed to rule out noise artifacts and separate signals from background, a good rule of thumb for 
detectability, which we follow here, is to consider a single detector and impose a matched-filtering 
signal-to-noise ratio (SNR) threshold of $8$ as a proxy for detectability by the network \citep{ratesdoc}.

The detection rate $R_d$ can be expressed in terms of the merger rate per unit comoving volume per unit time ${\cal R}$ as an integral over all possible merger redshifts
\begin{equation}\label{Rd}
R_d = \int_0^\infty {\cal R} (z) \frac{dV_c}{dz} \frac{dt_s}{dt_o} (z) f_d(z)  dz\, .
\end{equation}
Here the comoving volume of a given redshift slice $dV_c/dz$ is computed (e.g., see \citealt{1999astro.ph..5116H}) 
by assuming WMAP9 parameters $\Omega_M = 0.282$, $\Omega_\Lambda=0.718$,  $h=0.697$ \citep{2012arXiv1212.5226H}.  The factor $dt_s / dt_o = 1/(1+z)$ is the ratio between the clock at the source redshift $z$, which measures the merger rate, and a clock at the Earth, which measures the detection rate.  
The detection fraction $f_d(z)$ measures the fraction of sources of a given mass at a given redshift that will be detectable.

The expectation value of the optimal matched filtering SNR for an overhead, face-on source is given by 
\begin{equation}
\textrm{SNR}_{\textrm{opt}}^2 = 4 \Re \int_0^\infty
\frac{|\tilde{h}(f)|^2}{S_n(f)} df\, ,
\end{equation}
where $S_n(f)$ is the noise power spectral density of the detector and $\tilde{h}(f)$ is the frequency-domain 
representation of a GW signal. For the waveforms $h(t)$, we use spinning, non-precessing, inspiral-merger-ringdown 
effective-one-body waveforms calibrated to numerical relativity waveforms~\citep{Taracchini:2012}, specifically 
their {\it SEOBNRv1} implementation in the LIGO Scientific Collaboration Algorithm Library \citep{LAL}. The waveform is corrected 
for the redshift $z$ and its amplitude is inversely proportional to the luminosity distance to the source, 
which scales with $z$ according to the assumed standard cosmology \citep{1999astro.ph..5116H}.  
Face-on, overhead equal-mass binaries with a total rest-frame mass of $200 M_\odot$ can be detected to a redshift 
of $z\sim2$ (luminosity distance of $\sim 16$ Gpc) if the components are non-spinning, or $z\sim2.6$ (luminosity 
distance of $\sim 21$ Gpc) if both components have aligned spins with a dimensionless magnitude of $0.6$.

The actual SNR 
for a binary with a given sky location $\alpha, \delta$, inclination $\iota$, and polarization $\psi$ is given 
by $\textrm{SNR}_{\textrm{opt}} \times \Theta(\alpha,\delta,\iota,\psi)/4$, where the projection function $\Theta$
is defined in \cite{Finn:1996}. We compute the fraction $f_d(z)$ by taking into account the (numerically computed) 
cumulative distribution function (CDF) of $\Theta/4$, and setting the detectability threshold at an SNR of 8 as 
follows \citep[e.g.,][]{CygnusX3:2012}:
\begin{equation}
f_d (z) = 1 - \textrm{CDF}_{\Theta/4} \left[\min\left(\frac{8}{\textrm{SNR}_{\textrm{opt}}(z)},1\right)\right] \, .
\end{equation}

Under the simplifying assumption that the merger rate $\cal R$ is constant in redshift, which we use here, 
Eq.~(\ref{Rd}) for the detection rate can be written as
\begin{equation}\label{Rdsimple}
R_d = {\cal R} \overline{V}_c\, ,
\end{equation}
where we have defined the detection-weighted sensitive comoving volume $\overline{V}_c$ for a given rest-frame mass combination as
\begin{equation}
\label{DWSCV}
\overline{V}_c = \int_0^\infty \frac{dV_c}{dz} f_d(z) \frac{1}{1+z} dz\, .
\end{equation}

\begin{figure}
\begin{center}
\includegraphics*[width=0.5\textwidth]{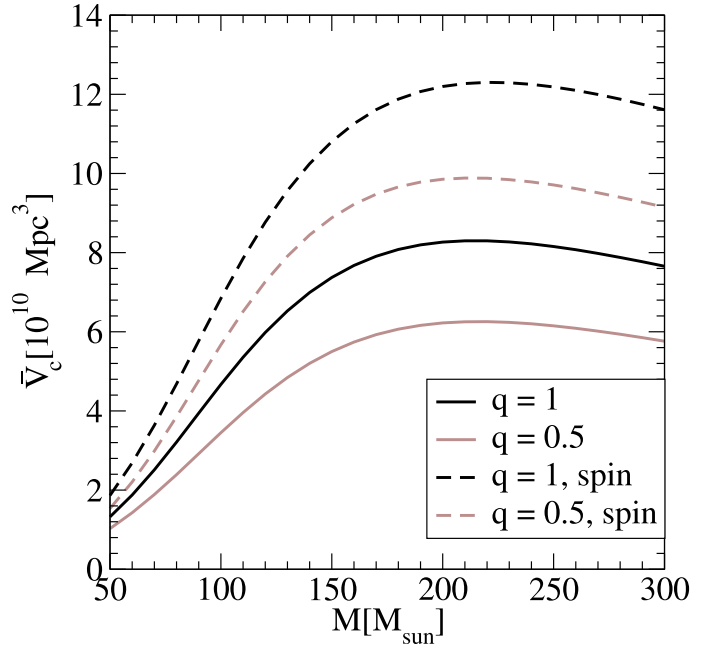}
\caption{The detection-weighted sensitive comoving volume $\overline{V}_c$ within 
which a coalescence could be detected for BH binaries with total source-frame mass $M$, 
mass ratios $q = 1$ or $0.5$, either non-spinning or with dimensionless 
spins magnitudes $0.6$.}
\label{Fig:Vc} 
\end{center}
\end{figure}

In Fig.~\ref{Fig:Vc} we plot the detection-weighted comoving volume as a function of the total system mass 
in the source frame for mass ratios $q=0.5$ and $q=1$, and spins that are either zero or both 
aligned with the angular momentum, with dimensionless magnitude $0.6$.  We use the noise spectral density of 
the high-power, zero-detuning configuration \citep{PSD:AL}, which corresponds to nominal 
sensitivity that is expected for advanced LIGO detectors by the end of the decade \citep{scenarios}. 
As expected, we find that when the mass ratio is 
equal and the aligned spins are high, signals contain more power and the sensitive volume is increased.

Equation~(\ref{Rdsimple}) and Fig.~\ref{Fig:Vc} allow us to convert a merger rate into a detection rate. 
In Section~\ref{sec:popsyn} we obtained a most optimistic merger rate of $5 \times 10^{-9}\,{\rm Mpc}^{-3}\,{\rm yr}^{-1}$ for BH-BH mergers with total mass above $100\msun$. The peak of the total mass distribution for this model appears at $200\msun$ and the BHs are of comparable mass. The weighted detectable volume for $M_{\rm tot}=200\msun$ is 
$6$--$12 \times 10^{10}$ Mpc$^3$ 
depending on BH spins, which yields $300$--$600$ 
detections per year. Our more realistic estimate, which accounts for pair-instability SNe, produces
a merger rate estimate of very massive BH-BH binaries of $5 \times 10^{-11}\,{\rm Mpc}^{-3}\,{\rm yr}^{-1}$, 
with a predicted detection rate of $3$--$6$ a year. If we account for potential uncertainties in CE evolution 
the merger rate decreases to zero, and no detections are predicted for very massive BH-BH binaries in the isolated 
evolutionary scenario (Model 1).

In Section~\ref{sec:OoM} we argued for a plausible order-of-magnitude
formation rate of $\sim 10^{-8}\,{\rm Mpc}^{-3}\,{\rm yr}^{-1}$ for
wide binaries with $100\msun$ or more massive BHs. In
Section~\ref{sec:dynamics} we estimated that $\sim 10\%$ of these wide
binaries may merge within a Hubble time due to dynamical interactions
with other stars. This yields a merger rate of $\sim 10^{-9}\,{\rm
  Mpc}^{-3}\,{\rm yr}^{-1}$ and a corresponding detection rate of
$\sim 100$ per year if dynamical interactions
are efficient in forming close very massive BH-BH binaries (Model
2). 

We also considered intermediate-mass-ratio binaries of 
$\gtrsim 100\,M_\odot$ BHs with $\sim 10\,M_\odot$ BH companions, and concluded that a formation rate of $\sim 10^{-9}\,{\rm Mpc}^{-3}\,{\rm yr}^{-1}$ was plausible from order-of-magnitude arguments.  Assuming that such systems can be brought to merge within a Hubble time by efficient dynamical hardening of the binary \citep[e.g.,][]{Mandel:2008}, they can be detected to $z\sim 0.5$,
  which decreases the sensitive volume by a factor of $\sim 20$
  relative to the $100\,M_\odot$--$100\,M_\odot$ case discussed
  above.  Therefore, up to $\sim 5$ intermediate-mass-ratio systems formed from very massive stars with lower-mass companions could be detected per year with advanced GW
  instruments.

It is also interesting to compare the population synthesis model predictions with existing upper 
limits from initial LIGO and Virgo runs.  Their sensitivity to massive BH binaries 
was poor because of the high value of the low-frequency cutoff, so the more interesting 
existing LIGO/Virgo constraint will come from BH binaries with total mass between $50$ and $100$ solar masses.
As reported in Section~\ref{sec:popsyn}, such binaries represent $93\%$ of our entire close BH-BH 
population, and this corresponds to a merger rate of $7 \times 10^{-8}\,{\rm Mpc}^{-3}\,{\rm yr}^{-1}$ 
for our most optimistic prediction. These binaries have formed from stars with initial mass
$M_{\rm zams} \approx 150-500\msun$. 
The initial LIGO/Virgo upper limits are $1.7 \times 10^{-7}\,{\rm Mpc}^{-3}\,{\rm yr}^{-1}$ and  
$0.9 \times 10^{-7}\,{\rm Mpc}^{-3}\,{\rm yr}^{-1}$ for equal mass BH binaries
in the mass bins 
$M_{\rm tot}=54$--$73\msun$ and $M_{\rm tot}=73$--$91\msun$, respectively \citep{2013PhRvD..87b2002A}.
Thus even our most optimistic rates do not violate the current upper limits for BH-BH detection.  

Given our predictions of preferentially aligned spins for isolated binaries and the possibility of spins 
misaligned with the orbital angular momentum for systems in which dynamics played an important role, it 
is interesting to ask how well such systems can be distinguished based on gravitational-wave observations.   
Several studies that considered a few individual events suggest that at least the spin magnitude of the 
more massive component and the angle between it and the orbital angular momentum could be measured, 
especially if this component is rapidly spinning, although it may prove difficult to measure the spin of 
the lower-mass secondary \citep[e.g.,][]{VanderSluys:2008b, Raymond:2010}. A larger study \citep{Vitale:2014} 
supports this view, and highlights the improvements to spin measurement when the binary is nearly edge-on 
rather than face-on, and when the angular momentum of the secondary is small, reducing spin-spin terms.   
Furthermore, Bayesian model selection may allow us to determine whether there's significant support for 
precession induced  by spin-orbit misalignment even if individual parameters cannot be measured precisely 
\citep[e.g.][]{S6PE}.  Unfortunately, the large size of the parameter space makes difficult a fully 
systematic study based on astrophysically-motivated source distributions.

We conclude this section by pointing out the importance of {\it (i)} 
accurate analytical and numerical modeling of inspiral-merger-ringdown waveforms, 
since for massive BH-BH binaries, the SNR accumulates mostly 
during the last stages of the binary evolution (e.g., 
see \cite{2013CQGra..31b5012H}), and 
{\it (ii)} low-frequency detector sensitivity in searches for massive BH
binaries, which contribute most of the SNR below $\sim 35$ Hz.  The sensitive volume plotted in Fig.~\ref{Fig:Vc} assumes that the predicted high-power, zero-detuning sensitivity \citep{PSD:AL} will be achieved down to $10$ Hz, which will require significant commissioning effort.  Meanwhile, 
third-generation ground-based detectors such as the
Einstein Telescope \citep{ET}, with sensitivity down to a few Hz, will
be able to probe such massive BH-BH binaries to $z\sim 15$
\cite[e.g.,][]{2011GReGr..43..485G}.

\section{Discussion and conclusions}

Since GWs from massive BH binaries can be detected to
cosmological distances, we have explored the event rates for the mergers of
these systems. 
We find that only low-metallicity environments ($Z\ltorder 0.1$--$0.4\,Z_\odot$) may be favorable for 
the formation of very massive stellar-origin BHs with mass exceeding $100\msun$. 
The formation of such BHs is possible if
{\it (i)} the initial mass function (IMF) extends above $500\msun$, {\it (ii)} pair-instability SNe do not disrupt all 
stars above $500\msun$, and {\it (iii)} stellar winds for such massive stars are not greatly 
underestimated. The formation of close massive BH-BH binaries  requires that after the main sequence {\it
  (iv)} very massive stars above $500\msun$  
expand significantly (by more than a factor of $2$), {\it (v)} their H-rich envelopes have a mass 
larger than $10$--$100\msun$, {\it (vi)} the evolution of such a binary involves a common envelope 
phase, and {\it (vii)} the binary can survive the common envelope phase while the donor star is 
a very massive Hertzsprung gap star.  If conditions {\it (iv)} through {\it (vii)} are not met, then 
isolated binary evolution (i.e., field stellar populations) may produce only wide massive BH-BH 
binaries. We point out that even if these requirements are not met, there are several dynamical 
processes that could lead to efficient lowering of the coalescence time of wide massive BH-BH
binaries both in dense stellar environments (cluster binary-single interactions) and in low-density field populations (Kozai mechanism in triple systems). 

The resulting BH-BH merger rates depend sensitively on the amount of star forming mass with 
low metallicity at redshifts $z<2$ (the maximum distance at which a $100$--$100\msun$ BH-BH 
binary will be detectable with the advanced LIGO/Virgo network). The amount of low metallicity 
star formation in the last Gyr may have been as high as $\sim 50\%$ of total star formation 
\citep{2008MNRAS.391.1117P}, and may have been even higher at the redshifts $z>1$ that 
dominate our overall rates. Population synthesis models predict that $75\%$ of the close massive BH-BH systems merge within 1 Gyr of formation. 
Based on simple estimates we find that realistic advanced LIGO/Virgo detection
rates for these massive BH-BH systems are on the order of a few per year. However, the large 
uncertainties that burden our predictions allow for rates as high as hundreds of detections 
per year to as low as no detections at all. 

BH-BH systems originating from isolated binaries of very massive stars would likely have rather large 
mass ratios ($q \gtorder 0.8$) and aligned spins.   For a core collapse of a massive star that was 
spin-aligned with the orbit via tidal interactions with its companion, and that ejects no mass and 
has no linear momentum kick resulting from the collapse, our strong expectation is that the compact 
remnant would have a spin aligned with the orbit. Given our still-rudimentary understanding of core 
collapses there are of course situations in which this might not be accurate (e.g., oppositely-directed 
neutrino jets on opposite sides of the proto-neutron star could impart spin angular momentum without 
imparting linear momentum), but at present such scenarios seem contrived. Regular ($M_{\rm zams} < 150 \msun$) 
stars could not produce binaries with total mass in the $M_{\rm tot} \gtrsim 100$--$200 \msun$ range.  
Therefore, detections of coalescences between massive, aligned, rapidly spinning BHs would uniquely 
identify systems originating from isolated binaries of very massive stars.  Such detections would 
indicate that the stellar IMF extends well beyond previously considered limits, and probably 
as high as  $M_{\rm zams} >  500 \msun$, and that massive progenitor binaries survive 
common envelope events even if the donors initiating these events are still in early 
evolutionary stages (i.e., on Hertzsprung gap). This would also argue against models of very 
massive stars that predict minimal expansion during post main-sequence evolution.

If observations show evidence for significant misalignment of BH spins, then either these
systems received kicks via asymmetric neutrino emission or dynamical processes must have been 
involved in the formation of the merging BH-BH system.  Moreover, \citet{Kalogera:2000} argued 
that significant spin--orbit misalignment is unlikely even with supernova kicks, unless dynamical 
effects are involved.   This latter scenario would support the first set of published models of 
very massive stars that show no (or very small) radial expansion \citep{2013MNRAS.433.1114Y}.  
Alternatively, mergers of intermediate-mass BHs could be a consequence of dynamical processes in 
globular clusters or involving globular cluster mergers 
\citep{Fregeau:2006,AmaroSeoaneSantamaria:2009,2011GReGr..43..485G}. 
Regardless, the detection of such systems would yield valuable information 
about dynamical encounters and/or the multiplicity of very massive stars.  

If instead we do not detect any massive BH-BH binaries with advanced LIGO/Virgo, any or all 
of the effects {\it (i)--(vii)} discussed above may have contributed.

When mass is transferred from the secondary to the primary after the primary 
has evolved to become a BH, the binary may be visible as an 
ultra-luminous X-ray source (ULX), as the mass loss rate from the massive secondary can 
be very high.  This stage is likely to be quite short, given that both companions are 
assumed to be massive and will have quite similar lifetimes. If 10\% of all such binaries 
have a ULX stage lasting 1 million years---or, equivalently, all binaries have a ULX 
stage lasting 100,000 years---the space density of observable ULXs created through this 
channel will be equal to the merger rate times $10^5$ years.  Since this is only one of 
several channels for creating ULXs, an observed ULX space density of a few
$\times 10^{-2}$ Mpc$^{-3}$ \citep{Swartz:2011} yields an upper limit on the massive 
binary merger rate of a few $\times 10^{-7}$  Mpc$^{-3}$ yr$^{-1}$.  This is a factor of 
several hundred higher than our very rough estimate of $10^{-9}$ Mpc$^{-3}$ yr$^{-1}$, so 
it would be easy to hide the population of interest among the observed ULXs. 
   
One possible constraint on the population of massive stars comes from
rates of pair-instability supernovae.  If an IMF slope of $\sim -2.5$
is extended to arbitrarily high masses, then $\sim1\%$ of all stars
having a ZAMS mass above $8 M_\odot$ will have a ZAMS mass above $200
M_\odot$.  Hence, if a significant fraction of very massive stars end
their life in pair-instability supernovae, one might expect as many as
1\% of all core-collapse supernovae to be pair-instability
supernovae.  Meanwhile, recent work by \cite{Nicholl13} finds that this
fraction is no more than $10^{-4}$, and may be $< 10^{-5}$,
if the so-called ``superluminous'' supernovae are inconsistent with
expectations of pair instability supernovae \citep{Nicholl13}.  If
pair-instability supernovae only produce superluminous supernovae,
these observations constrain either the number of massive stars
exceeding $\sim 150\,M_\odot$ or the mass range of stars that produce
pair-instability supernovae.  In such a scenario, if advanced LIGO/Virgo observes a
large fraction of massive binary BH mergers, it would place
constraints on the pair-instability mechanism and the mass range for
which it occurs.  Unfortunately, pair-instability supernovae seem to
be able to produce a wide range of light
curves \citep{Kasen11,Whalen14} and until these light-curves can be
better understood, it will be difficult to make any firm conclusions
using superluminous supernova observations.

In conclusion, advanced LIGO/Virgo detectors are
sensitive to the merger of massive BH binaries out to extraordinary distances
($z\sim2$). We argue that the rate density of these mergers is
such that event rates of a few per year, and perhaps as many as hundreds per
year, are possible given current uncertainties in stellar evolutionary
physics. The upper limits or detections expected from the coming generation of
advanced ground-based GW instruments will provide unique insights into the
evolution of very massive stars.

\acknowledgements
The authors thank Mirek Giersz, Duncan Brown, and especially Vicky Kalogera for useful comments.  
KB and MW acknowledge support from Polish Science Foundation "Master2013" 
Subsidy, by Polish NCN grant SONATA BIS 2. KB also acknowledges
NASA Grant Number NNX09AV06A and NSF Grant Number HRD 1242090 awarded to the 
Center for Gravitational Wave Astronomy, UTB. AB acknowledges  
support from NSF Grant No. PHY-1208881 and NASA Grant NNX12AN10G.
DEH acknowledges support from NSF CAREER grant PHY-1151836.
IM was partly supported by STFC, including an ET R\&D grant funded within the ASPERA-2 framework.  
MCM acknowledges NASA grant NNX12AG29G, and a grant from the Simons Foundation 
(grant number 230349). MCM thanks the Department of Physics and Astronomy 
at Johns Hopkins University for hosting him during his sabbatical.
The authors thank the Kavli Institute for Theoretical Physics 
(supported by the NSF Grant No.~PHY11-25915) for hospitality 
during the genesis of this project.  
The authors acknowledge the Texas Advanced Computing Center (TACC) at The
University of Texas at Austin for providing computational resources used for
this study. 

\bibliographystyle{hapj}
\bibliography{massive}

\end{document}